\documentclass[sigplan,nonacm]{acmart}
\setcopyright{none}
\acmDOI{}
\acmISBN{}
\acmConference{}{}{}

\usepackage{microtype}
\usepackage{graphicx}
\usepackage{booktabs}
\usepackage{array}
\usepackage{amsmath}
\usepackage{xspace}
\usepackage{adjustbox}
\usepackage{pifont}
\usepackage{placeins}
\usepackage{algorithm}
\usepackage{algorithmic}
\usepackage{xurl}

\definecolor{mydarkblue}{rgb}{0,0.08,0.45}
\hypersetup{
  keeppdfinfo=true,
  pdftitle={RouteBalance: Fused Model Routing and Load Balancing for Heterogeneous LLM Serving},
  pdfauthor={Wei Da and Evangelia Kalyvianaki},
  pdfsubject={LLM serving, model routing, and load balancing},
  pdfkeywords={LLM serving, model routing, load balancing, scheduling},
  pdfborder={0 0 0}, pdfpagemode=UseNone,
  colorlinks=true, linkcolor=mydarkblue, citecolor=mydarkblue,
  filecolor=mydarkblue, urlcolor=mydarkblue, pdfview=FitH}

\newcommand{\sys}{RouteBalance\xspace}
\newcommand{\Circled}[1]{\ding{\numexpr171+#1\relax}}
\title[RouteBalance: Fused Model Routing and Load Balancing]{RouteBalance: Fused Model Routing and Load Balancing for Heterogeneous LLM Serving}
\author{Wei Da}
\affiliation{\institution{University of Cambridge}\country{United Kingdom}}
\author{Evangelia Kalyvianaki}
\affiliation{\institution{University of Cambridge}\country{United Kingdom}}

\date{}

\begin{document}
\begin{abstract}
Current heterogeneous LLM serving clusters decouple request scheduling into two independent tiers: learned model routers select a model based on the prompt without visibility into instance-level load, while load balancers assign requests to an instance of that chosen model without cross-model quality or cost awareness.
This decoupled hierarchy induces conflicting objectives and severely restricts the global optimization space. For example, a quality-oriented router may stubbornly direct 55\% of requests to a 72B model regardless of growing queues, inflating mean latency by 2.4$\times$ under high load. Conversely, a cost-minimizing router over-allocates traffic to the cheapest model, leaving capable instances underutilized and capping aggregate cluster throughput.
This paper presents \sys, a serving scheduler that unifies model routing and instance dispatching into a single decision. \sys dynamically batches incoming requests into short, load-adaptive rounds. In each round, it amortizes quality and response-length estimation across candidate models, takes a single snapshot of all instance loads, and solves a normalized assignment problem balancing quality, cost, and estimated latency.
On a 28-GPU heterogeneous cluster spanning four Qwen model sizes and diverse GPU architectures, \sys Pareto-dominates 231 of 246 baseline configurations across quality, latency, and cost simultaneously. These baselines comprise combinations of two learned routers with four downstream dispatchers, as well as five standalone dispatchers. At iso-quality, per-rate median reductions across baselines are 23--60\% in mean E2E latency and 26--37\% in serving cost. Code: \url{https://github.com/AKafakA/route-balance}.
\end{abstract}
\maketitle
\raggedbottom

\hypersetup{
  pdftitle={RouteBalance: Fused Model Routing and Load Balancing for Heterogeneous LLM Serving},
  pdfauthor={Wei Da and Evangelia Kalyvianaki},
  pdfsubject={LLM serving, model routing, and load balancing},
  pdfkeywords={LLM serving, model routing, load balancing, scheduling},
  linkcolor=mydarkblue, citecolor=mydarkblue,
  filecolor=mydarkblue, urlcolor=mydarkblue}

\section{Introduction}\label{sec:intro}
Production LLM serving clusters rarely run a single model on uniform hardware. To serve workloads that range from simple factual lookups to multi-step coding and reasoning, operators deploy fleets of models spanning an order of magnitude in parameter count across multiple GPU generations. A 3B or 7B model can answer routine prompts quickly at low price, while a 72B model handles harder prompts at higher cost and memory footprint. In a heterogeneous cluster, however, the trade-off between answer quality, latency, and price is not fixed by model size alone. It shifts continuously with hardware placement and live engine load~\cite{kwon2023vllm}. Existing serving stacks split this scheduling decision into two separate phases that operate with partial views of the cluster. First, a learned model router inspects the prompt and selects a target model based on predicted quality, occasionally factoring in token price ~\cite{best_route, ong2024routellm, avengers, chen2023frugalgpt, nvidiaLlmRouter}, without visibility into serving backend replica status and load. Second, a downstream load balancer or request scheduler picks a replica of that chosen model~\cite{githubGitHubLlmdllmd, da2026astrolabe, aibrix, dynamo}, without the ability to redirect the request to a different model. Under light traffic, this separation could work well enough. However, under load, the two phases work against each other. A quality-oriented router continues to send a fixed fraction of traffic to the large model regardless of its replicas' load. Conversely, a cost-minimizing router packs requests onto the smallest models, starving capable multi-GPU instances and bottlenecking cluster throughput. Once the routing decision is made, the downstream load balancer is constrained to dispatch requests only among replicas of the pre-selected model, forfeiting global optimization opportunities, if other models' replicas are underutilized.

We present \emph{\sys}, a serving scheduler that jointly selects the model and the instance for each incoming request in a single decision. To amortize overhead and enable coordinated dispatching, \sys buffers arrivals into short, load-adaptive \emph{scheduling rounds}, termed ``rounds'' to distinguish them from the execution batches formed by backend engines. Backend engines run unmodified vLLM servers with standard continuous batching, and each assigned request is dispatched immediately without waiting for the round to finish. For every round, \sys performs a single batched $k$-NN search over labeled prompts to jointly estimate each model's answer quality and output length. It queries instance loads once at the beginning of each round and maintains exact in-flight request counts locally via dead reckoning (\S\ref{sec:load}), projecting latency from expected output lengths and pre-profiled TPOT. Finally, \sys schedules requests in longest-first order, greedily assigning each request to the (model, instance) pair that maximizes a joint weighted score of quality, cost, and latency, updating instance state sequentially after each assignment. Deciding a round's requests together also allows cluster load to shift requests across models. \sys normalizes each request's quality gaps against the dynamic quality range of the entire round. Under heavier load, rounds naturally grow and cover a broader quality range, which shrinks small quality gaps and lets the latency estimate drive the dispatching decision. In contrast, when scores are normalized per request, quality gaps span the entire [0, 1] range regardless of queue depth, rendering model selection insensitive to load. At 80/10/10 (quality, latency, and cost weights), the 72B model's share stays flat at 48\% across all loads, whereas round-wide normalization drops its share from 39\% at 12 req/s to 32\% at 42 req/s (\S\ref{sec:split}, \S\ref{sec:ablation}).

On a 28-GPU cluster of A30, V100, and A100 GPUs serving Qwen2.5 3B to 72B, \sys at its eight weight presets beats 231 of the 246 baseline settings on quality, mean end-to-end (E2E) latency, and cost together. At equal quality, it lowers mean E2E latency by a median of 37\% and cost by 32\%, and the latency reduction grows with the load, from 23\% at 12 req/s to 60\% at 42 req/s. Two variants with the same estimator and weights isolate the decision gains. Compared to the split decision (RB-decoupled, \S\ref{sec:impl}), \sys is faster in 43 of the 48 preset--rate pairs (by up to 43\%) and cheaper in 42 (by up to 26\%). It is also faster in 47 pairs than the same decision made one request at a time (RB-single), which begins queuing requests at 24 req/s. Deciding a round at a time also keeps a request's wait in \sys before dispatch at 123--198 ms (\S\ref{sec:ablation}). Finally, with routing settings fixed on Qwen, no baseline setting beats \sys on Gemma~3.

This paper makes three contributions. \emph{First}, we show that decoupling model routing from load balancing inherently degrades latency and cost under load (\S\ref{sec:background}, \S\ref{sec:main}, \S\ref{sec:split}). \emph{Second}, we design and implement \sys, which unifies model routing and instance placement via load-adaptive rounds, batched estimation, dead-reckoned tracking, and round-wide normalization (\S\ref{sec:design}, \S\ref{sec:impl}). \emph{Third}, extensive cluster evaluations on Qwen2.5 and Gemma~3 demonstrate that \sys consistently outperforms existing routers and dispatchers, while targeted ablations isolate the gains of each core design decision. We further validate its generality under bursty traffic and budget filtering (\S\ref{sec:eval}).

\section{Background and Related Work}\label{sec:background}

\textbf{Heterogeneous LLM serving.} An LLM request runs in two phases: \emph{prefill} processes the prompt and populates the KV cache, largely determining the time to first token (TTFT), while \emph{decode} generates tokens autoregressively at a time per output token (TPOT) that dominates long responses. Engines such as vLLM~\cite{kwon2023vllm} employ \emph{continuous batching}, where finished requests exit and waiting requests join the running batch at each step up to the batch capacity limit. Consequently, TPOT grows with the batch size, and requests arriving at a full engine must wait for free slots in running batch. This dynamic is further complicated by hardware and model heterogeneity. Although a larger model typically costs more per token and handles complex prompts better, it can outpace a smaller model depending on the underlying hardware. For example, in our fleet, a 14B model on four V100s has a lower median TPOT than a 7B model on one A30 (Table~\ref{tab:cluster}).

\begin{table*}[t]
\centering\footnotesize\setlength{\tabcolsep}{2.5pt}
\caption{Comparison of routing and serving systems across scopes and input signals. Decisions are made per request window (\sys, SBS), per request group (QLM), or per individual request (all others). \textbf{M}: model selection. \textbf{I}: instance selection across replicas (\textsuperscript{*}single instance per model in evaluation; model choice fixes the instance). \textbf{S}: request re-ordering rather than strict FCFS (\textsuperscript{\textdegree}opt-in priority ordering rather than performance optimization). \textbf{GPUs}: evaluated on heterogeneous GPU tiers. \textbf{\$}: explicitly weighs model token prices.}
\label{tab:comparison}
\begin{tabular}{>{\raggedright\arraybackslash}p{5.5cm}>{\raggedright\arraybackslash}p{7.9cm}ccccc}
\toprule
System & Signals & M & I & S & GPUs & \$ \\
\midrule
RouteLLM~\cite{ong2024routellm} & estimated win rate between strong and weak models & \ding{51} & & & & \\
AvengersPro~\cite{avengers} & nearest embedding cluster accuracy and cost estimates & \ding{51} & & & & \ding{51} \\
BEST-Route~\cite{best_route} & predicted model match score, sample count & \ding{51} & & & & \ding{51} \\
NVIDIA LLM Router~\cite{nvidiaLlmRouter} & intent match or domain classification score & \ding{51} & & & & \ding{51} \\
\midrule
vLLM Semantic Router~\cite{semantic_router} & per-model benchmark score, observed latency, in-flight counts & \ding{51} & & & & \ding{51} \\
HW-Router~\cite{hwrouter} & predicted quality, per-model queues, KV-cache, TTFT/TPOT & \ding{51} & \ding{51}\textsuperscript{*} & & & \\
BEAR~\cite{patel2026minddelaylatencyawarequery} & predicted accuracy, BEA-simulated TTFT & \ding{51} & \ding{51}\textsuperscript{*} & & & \ding{51} \\
\midrule
llm-d endpoint picker~\cite{githubGitHubLlmdllmd} & queue and KV-cache gauges (50 ms poll), prefix match rate & & \ding{51} & \ding{51}\textsuperscript{\textdegree} & & \\
AIBrix gateway~\cite{aibrix} & instance metrics, in-flight counts, token rate, prefix match rate & & \ding{51} & \ding{51}\textsuperscript{\textdegree} & \ding{51} & \\
NVIDIA Dynamo router~\cite{dynamo} & cached KV blocks, router-tracked active blocks & & \ding{51} & \ding{51}\textsuperscript{\textdegree} & & \\
SGLang router~\cite{sglangGateway} & prefix match rate (via local prefix tree), in-flight counts & & \ding{51} & & & \\
Llumnix~\cite{llumnix} & KV-cache freeness with queued demand, request priority & & \ding{51} & \ding{51}\textsuperscript{\textdegree} & & \\
Astrolabe~\cite{da2026astrolabe} & estimated length, simulated latency of two sampled instances & & \ding{51} & & & \\
\midrule
QLM~\cite{qlm} & estimated group wait time, model-swap overhead & & \ding{51} & \ding{51} & \ding{51} & \\
PolyServe~\cite{shi2025polyserveefficientmultislo} & iteration times, server wait time, nearest deadline & & \ding{51} & \ding{51} & & \\
SBS~\cite{sbs2025} & in-flight counts, queued tokens, batch and KV-cache size & & \ding{51} & \ding{51} & & \\
\midrule
\sys & predicted quality and length, profiled TPOT, instance load & \ding{51} & \ding{51} & \ding{51} & \ding{51} & \ding{51} \\
\bottomrule
\end{tabular}
\end{table*}

\textbf{Model routing.} Cascades and trained routers select a model for each prompt based on its complexity~\cite{chen2023frugalgpt, ong2024routellm, feng2024graphrouter, jitkrittum2025universal, hu2024routerbench}, including AvengersPro~\cite{avengers} and BEST-Route~\cite{best_route}, the two open-source routers we use as baselines. Typically, such routers only select the target model, leaving a separate load balancer to forward the dispatched request to a specific instance within that model's cluster. Two concurrent works, HW-Router~\cite{hwrouter} and BEAR~\cite{patel2026minddelaylatencyawarequery}, incorporate latency estimates into model selection. Both make per-request decisions and were evaluated with a single instance per model, where choosing a model fixes the instance and obviates the need for load balancing. HW-Router trades off quality against latency rather than price. Moreover, because its engine queue lengths are updated via periodic polling, they omit in-flight requests dispatched since the last poll. BEAR uses its BEA estimator to simulate TTFT from scheduler snapshots: its primary policy maximizes predicted accuracy minus cost among instances meeting a preset TTFT target, while a second variant directly penalizes TTFT instead of enforcing hard constraints. The vLLM Semantic Router supports multiple replicas, but its balancing policy tracks load only at the model level (Appendix~\ref{app:vsr}), delegating instance-level balancing to Envoy (e.g., via round-robin) and explicitly marking per-instance scheduling as out of scope ~\cite{vllmsr_scope}. In contrast, \sys jointly scores quality, cost, and E2E latency across every individual instance (Eq.~\ref{eq:objective}).

\textbf{Load balancing and request scheduling.} As discussed above, existing load balancers and schedulers take the target model as given. The \mbox{llm-d} endpoint picker~\cite{githubGitHubLlmdllmd}, the AIBrix gateway~\cite{aibrix}, and other load balancers~\cite{dynamo, sglangGateway, llumnix, da2026astrolabe} dispatch each request to a replica of its specified model, while request schedulers could also re-order waiting requests before dispatch~\cite{chen2025slosserveoptimizedservingmultislo, qlm, shi2025polyserveefficientmultislo, sbs2025, 11723265}. \sys, in contrast, jointly selects the model and the instance based on estimated quality, cost, and latency across models of varying sizes and GPU tiers. It also leaves each instance's engine untouched, including the local scheduler that batches pending requests into the running batch. Table~\ref{tab:comparison} summarizes these design dimensions, comparing \sys with existing model routers, load balancers, and schedulers.

\textbf{Deployment Planner.} Deployment planners such as HexGen~\cite{hexgen}, BOute~\cite{boute}, and ThunderServe~\cite{thunderserve} determine, on a longer timescale, how each model is partitioned and replicated across a heterogeneous cluster's GPUs. BOute integrates RouteLLM into its configuration search: for a fixed workload trace and a cluster simulator, it jointly optimizes routing thresholds and multi-model deployment to trade simulated latency against estimated quality under a cost budget. In contrast, \sys takes a pre-deployed fleet of models as given and dynamically dispatches each incoming request at runtime. Consequently, \sys is complementary to these deployment planners, and a change in underlying cluster placement simply presents a new model fleet, requiring no retraining of \sys's estimators.

\section{System Design}\label{sec:design}

\begin{figure*}[t]
\centering
\includegraphics[width=0.9\textwidth]{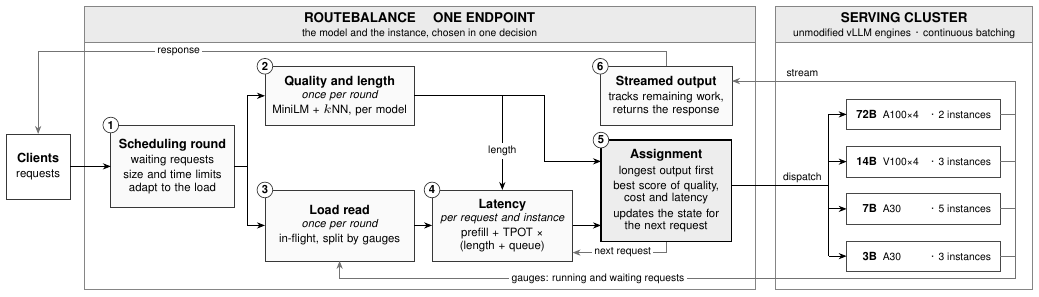}
\caption{\sys's request path, numbered by the steps of \S\ref{sec:design}.}
\label{fig:arch}
\Description{Diagram of RouteBalance's request path, from clients through scheduling and estimation to model instances.}
\end{figure*}

\sys sits between the clients and the serving cluster. Clients send ordinary generation requests to its one endpoint, and for each request \sys chooses a model and an instance of that model in one decision and forwards the request there (Figure~\ref{fig:arch}). 

A request waits in a queue until the scheduler opens a \emph{scheduling round} (\Circled{1}, \S \ref{sec:rounds}). For the round's requests, the estimator predicts each model's answer quality and output length from the prompt (\Circled{2}, \S \ref{sec:estimates}), and the scheduler reads every instance's load once (\Circled{3}, \S \ref{sec:load}). The scheduler then takes the requests one by one after sorting by the expected output length. For each request, it estimates the latency on each instance (\Circled{4}, \S \ref{sec:latency}), assigns the request to the instance with the best combined score, and counts the assignment in the state used for subsequent requests (\Circled{5}, \S \ref{sec:objective}). \sys then reads the streamed output to dynamically update its view of each instance's remaining workload, and returns the response to clients when it completes (\Circled{6}).

\subsection{Scheduling rounds}\label{sec:rounds}

A scheduling round is a short window in which \sys centrally buffers incoming requests and schedules them sequentially within a shared decision context. A round begins upon the arrival of the first pending request, and closes once it reaches its round-size limit or the timeout expires. Requests arriving while the current round is being scheduled are enqueued for the next round. To adapt to system load, \sys adjusts the round-capacity limit dynamically based on the fraction of instances whose engines report waiting requests: from 4 requests (50~ms window) below 0.3, 16 (100~ms) below 0.7, and 32 (200~ms) otherwise. When high arrival rates cause requests to accumulate at the scheduler, \sys eagerly ingests all currently buffered requests (up to 32) into the next round. This introduces few pipeline bubbles because the backend instances are already fully utilized, with no free slots to execute requests immediately. Conversely, forming a larger round amortizes overhead, as the k-NN embedding in \S\ref{sec:estimates} and load probing in \S \ref{sec:load}, and allows intra-round scheduling to coordinate assignments across more pending requests, effectively mitigating herd behavior, where concurrent requests are simultaneously assigned to the same instance.

\subsection{Quality and length estimates}\label{sec:estimates}

For every request and model, \sys needs the expected quality of the model's answer and its expected output length for the scoring in \S\ref{sec:objective}. We use a $k$-nearest-neighbor estimator over a labeled corpus of training prompts to estimate both together, keeping runtime serving overhead minimal. Each training prompt is embedded with MiniLM~\cite{wang2020minilm}, and each model's answer is recorded with its output length and quality score evaluated by an independent LLM judge (\S\ref{sec:setup}) to prevent self-evaluation bias. For a new request, the estimator finds its $k{=}50$ nearest training prompts by cosine similarity and returns the weighted average of their quality scores and output lengths for each model, with weights defined as $1/(d{+}10^{-6})$ for distance $d = 1 - \text{similarity}$. Because all requests in an incoming round are embedded and queried together, the estimator runs once per round to amortize this overhead across the entire cohort.

\subsection{Instance load and dead reckoning}\label{sec:load}

Tracking instance load at runtime faces two key challenges: polling remote endpoints per request could introduce accumulated network overhead, while engine gauges inherently lag behind the true instance workload, because vLLM updates them only upon completed execution steps. To address both issues, \sys employs \emph{dead reckoning}, a technique from distributed tracking systems where a node projects state locally from known events rather than stale remote queries.
When opening a round, \sys probes each instance's metrics endpoint only once to establish a baseline, amortizing the polling cost across all requests in the round. Because all traffic passes through \sys, the router also tracks the exact in-flight request count for each instance. It calibrates the baseline by bounding running requests to $\min(\text{engine gauge}, \text{in-flight})$ and treating any remaining in-flight requests as queued. As intra-round assignments proceed, \sys locally projects each instance's pending decode tokens by subtracting streamed tokens from each in-flight request's estimated length with no repeated polling and less exposure to stale gauges.

\subsection{Latency}\label{sec:latency}

Each tier $t$ (the instance type pairing a model with a GPU type) is characterized by a per-token decode time $\tau_t$, measured as the median TPOT with up to 96 concurrent requests, and a linear prefill time $P_t(p) = a_t + b_t p$ for a prompt of $p$ tokens (fitted as Table~\ref{tab:cluster} for $\tau_t$ and the coefficients $a_t, b_t$). Here, $a_t$ represents the fixed prefill initiation overhead and $b_t$ represents the per-prompt-token processing time. For a request with expected output length $\hat{L}$ on tier $t$ with $n$ running requests (capped at 256) and $D$ pending decode tokens, the latency estimate is

\begin{equation}
\ell = P_t(p) + \tau_t\,\bigl(\hat{L} + \delta\,D/\max(n,1)\bigr),
\label{eq:latency}
\end{equation}

This latency estimate accounts for the request's own prefill and decode stages, adding the queuing delay when routed to a busy instance. The queue term is enabled (\(\delta=1\)) when the tracked waiting count is positive or the running batch reaches its capacity, and \(0\) otherwise. Even on an idle instance ($\delta = 0$), the baseline latency term differentiates tiers by their hardware execution speed (Table~\ref{tab:cluster}). Within a round, dead reckoning updates the selected instance's waiting count and pending decode tokens after each assignment. \S\ref{sec:ablation} evaluates the contribution of the latency term to the scheduling objective and compares this analytical estimate against learned per-tier regressors fitted on instance states.

\subsection{The objective and the assignment}\label{sec:objective}

For request $r$ in round $R$ and candidate instance $i$ serving model $m(i)$, \sys scores
\begin{equation}
S_{r,i} = w_q\, Q_{r,i} + w_l\, T_{r,i} + w_c\, C_{r,i},\; w_q{+}w_l{+}w_c = 1,
\label{eq:objective}
\end{equation}
where every term lies in $[0,1]$ and higher is better (Table~\ref{tab:notation}):
\begin{equation}
\begin{aligned}
Q_{r,i} &= \frac{q_{r,m(i)} - q^{R}_{\min}}{q^{R}_{\max} - q^{R}_{\min} + \epsilon},\qquad
C_{r,i} = 1 - \frac{c_{r,i}}{c^{R}_{\max} + \epsilon},\\
T_{r,i} &= 1 - \frac{\ell_{r,i}}{\ell^{R}_{\max}(r) + \epsilon}.
\end{aligned}
\label{eq:terms}
\end{equation}
Here $\epsilon=10^{-8}$ is a numerical stabilizer, $q$ is the estimated quality, $c$ the estimated cost (the prompt and the $k$-NN expected output length at the tier's prices), and $\ell$ is the latency estimate from Eq.~\ref{eq:latency}. The quality range ($q^R_{\max} - q^R_{\min}$) and cost scale ($c^R_{\max}$) are computed across all requests and instances upfront, while $\ell^{R}_{\max}(r)$ dynamically tracks the rolling maximum latency evaluated up to request $r$. Because this range naturally widens as rounds expand under higher load, the tier mix dynamically adapts to traffic demands (Appendix~\ref{app:reqnorm}). The weight vector $w = (w_q, w_l, w_c)$ allows operators to trade off quality, latency, and cost.

\begin{table}[t]
\centering\small
\caption{Notation of Eqs.~\ref{eq:latency}--\ref{eq:terms}.}
\label{tab:notation}
\resizebox{\columnwidth}{!}{\begin{tabular}{ll}
\toprule
Symbol & Meaning \\
\midrule
$r$, $R$ & a request, the requests of its round \\
$i$, $I$, $m(i)$ & an instance, all instances, the model $i$ serves \\
$q_{r,m}$ & estimated quality of $m$'s answer to $r$, in $[0,1]$ \\
$\hat{L}_{r,m}$ & expected output tokens of $m$ on $r$ \\
$c_{r,i}$ & estimated cost: the prompt and $\hat{L}_{r,m(i)}$ at $i$'s prices \\
$\tau_t$, $P_t$ & tier $t$'s per-token decode time and prefill time \\
$n$, $D$ & an instance's running requests and pending decode tokens \\
$\ell_{r,i}$ & latency estimate of $r$ on $i$ as loaded now \\
$q^R_{\min}, q^R_{\max}, c^R_{\max}$ & the round's quality range and largest cost \\
$\ell^R_{\max}(r)$ & the largest latency estimate of the round so far \\
$w_q, w_l, w_c$ & weights of quality, latency and cost, summing to 1 \\
\bottomrule
\end{tabular}}
\end{table}

Algorithm~\ref{alg:round} shows the complete process within one round. Requests are taken in longest-processing-time-first (LPT) order by their largest expected length across models~\cite{graham1969bounds}. Then each request is scored on every instance from the current state and assigned to the highest-scoring candidate, updating the instance's state as described in \S\ref{sec:load} before scheduling next requests. Ties are broken first by favoring the instance with fewer active requests, and then in least-recently-used (LRU) order across instances. A round costs one estimator call, one load read, and $O(|R|\cdot|I|)$ scoring steps for $|R|$ requests and $|I|$ instances.

\begin{algorithm}[t]
\caption{One \sys scheduling round}\label{alg:round}
\begin{algorithmic}[1]
\REQUIRE round $R$, instances $I$, weights $(w_q, w_l, w_c)$
\STATE $q, \hat{L} \gets$ \textsc{Estimate}$(R)$ \COMMENT{one batched call}
\STATE $s \gets$ \textsc{ReadLoad}$(I)$ \COMMENT{one read, in-flight counts by dead reckoning}
\STATE $q^R_{\min}, q^R_{\max}, c^R_{\max} \gets$ over $R \times I$;\; $\ell^R_{\max} \gets 0$
\FORALL{$r \in R$ in LPT order of $\max_m \hat{L}_{r,m}$}
  \FORALL{$i \in I$}
    \STATE $\ell_{r,i} \gets$ Eq.~(\ref{eq:latency}) on $s_i$
  \ENDFOR
  \STATE $\ell^R_{\max} \gets \max(\ell^R_{\max}, \max_i \ell_{r,i})$;\; $S_{r,i} \gets$ Eq.~(\ref{eq:objective})
  \STATE $i^\star \gets \arg\max_i S_{r,i}$ \COMMENT{ties: queue, recency, order}
  \STATE $s_{i^\star} \gets s_{i^\star}$ + one waiting request and $\hat{L}_{r,m(i^\star)}$ tokens
  \STATE \textsc{Dispatch}$(r, i^\star)$
\ENDFOR
\end{algorithmic}
\end{algorithm}

\sys also supports budget-aware scheduling via an admission filter that retains only candidate instances whose expected cost falls within per-request budget. If no candidate qualifies, the filter falls back to those with the lowest worst-case cost. Appendix~\ref{app:budget} details this admission mechanism, and \S\ref{sec:generality} evaluates its effectiveness.

\section{Implementation}\label{sec:impl}

\sys is implemented as a Python service of approximately 3{,}300 lines of code, served via FastAPI on Uvicorn matching vLLM's server. An additional 900 lines implement the baselines within the same service. The backend engines are official, unmodified vLLM 0.18.1 servers. \sys queries their Prometheus metrics endpoints to monitor load and streams responses with chunk-level token usage to track decoding progress. The estimator uses MiniLM on the CPU with an exact FAISS inner-product index~\cite{faiss}, while scheduling rounds execute on a dedicated 4-thread pool off the main event loop to prevent stalling other active response streams.

Every baseline runs behind the same endpoint, in the same service, and across the identical set of engines and clients. For the routed baselines, a load-agnostic model router first selects the target model, and a downstream dispatcher then assigns an instance of that model. Neither HW-Router nor BEAR, the two concurrent load-aware model routers discussed in \S\ref{sec:background}, can run unmodified on our cluster as originally released to support multi-replica and multi-node serving: HW-Router assumes a single endpoint per model, while BEAR reads state snapshots directly from shared memory on the router host, whereas our 13 backend engines are distributed across distinct nodes. We compare \sys against the following baselines and two of its own variants:

\textbf{AvengersPro}~\cite{avengers} clusters prompt embeddings with $k$-means and routes each prompt to the model with the best performance--efficiency score in its cluster, which weighs normalized accuracy against normalized cost by $pw$. We run the authors' code with its default routing settings and sweep $pw \in \{0.25, 0.39, 0.53, 0.70, 0.80\}$.

\textbf{BEST-Route}~\cite{best_route} uses a DeBERTa-based ranker to estimate the probability that each candidate model's response matches or exceeds that of the strongest model. Each request is routed to the cheapest model whose score (that win-or-tie probability minus its complement) meets or exceeds a threshold $t$. The strongest model always qualifies by default. We then train and serve the authors' single-sample router using their original codebase, sweeping $t \in \{0, 0.3, 0.5, 0.7\}$.

\textbf{Dispatchers.} The dispatchers are least outstanding requests (LOR), round robin (RR), power-of-two choices (Po2), random, and lowest latency (LL). LOR picks the instance with the fewest reported in-flight requests (running plus waiting). Po2 samples two random instances and picks the one with fewer in-flights. LL picks the instance with the lowest mean per-token latency over its last 10 requests, following LiteLLM~\cite{litellm} rules. Each dispatcher runs either downstream of a router (assigning within the chosen model's instances) or standalone across all cluster instances.

\textbf{RB-decoupled} splits \sys's decision into two decoupled stages to mimic conventional router behavior: \sys's estimator and objective first choose the model (assuming an idle instance and normalizing per request), and LOR then places the request onto an instance.

\textbf{RB-single} runs \sys with a round size of one. Each request is scheduled individually across all instances using current load snapshots, as in HW-Router and BEAR. By keeping estimator, load tracker and scorer, it isolates the benefits of batching requests into scheduling rounds.

The routers run on their own pool of 32 threads, as many requests as \sys's largest round. The time a request waits for its router counts in its latency. Appendix~\ref{app:shipped} also runs
them one call at a time. LOR and Po2 read the engines' running and waiting counts
from a cache refreshed every 50 ms, matching the default refresh period of the \mbox{llm-d} endpoint picker~\cite{githubGitHubLlmdllmd}. Finally, the AvengersPro and BEST-Route artifacts are built for each model family from the same training prompts as \sys.
\section{Evaluation}\label{sec:eval}

\begin{figure*}[t]
\centering
\includegraphics[width=0.85\textwidth]{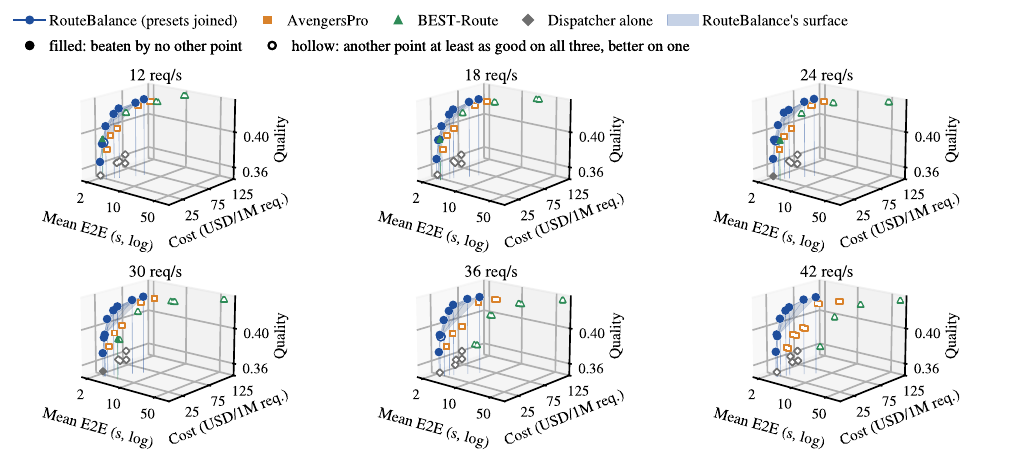}\\[2pt]
\includegraphics[width=0.85\textwidth]{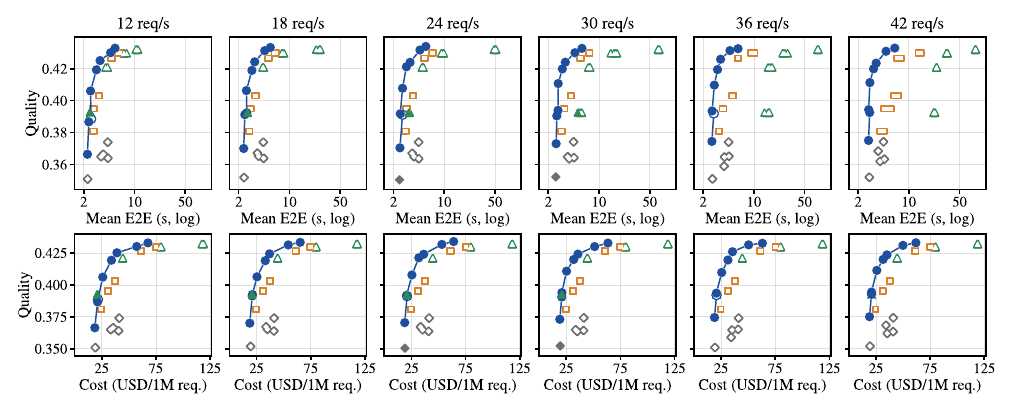}
\caption{Quality, mean E2E latency and cost of every \sys preset and baseline setting of the main comparison at each rate: in three dimensions (top, the surface interpolating \sys's eight presets), and as quality against mean E2E latency (log) and against cost (bottom, the line joining the presets). A filled point is nondominated in quality, mean E2E latency, and cost at the same arrival rate.}
\label{fig:frontier}
\Description{Plots comparing quality, mean end-to-end latency and serving cost for RouteBalance and baseline configurations across Qwen request rates.}
\end{figure*}
\subsection{Setup}\label{sec:setup}

\textbf{Testbed.} We run on CloudLab~\cite{cloudlab} with 13 instances across 28 GPUs (Table~\ref{tab:cluster}). Eight d7525 nodes each with one A30 (24 GB) serve Qwen2.5-3B and 7B, three c4130 nodes with four V100s (16 GB) serve the 14B, and two d8545 nodes with four A100s (40 GB) serve the 72B (both using 4-way tensor parallelism). \sys, all baseline schedulers, and the request client run on the CPUs of a single A100 node. For the second model family (\S\ref{sec:generality}), the A30 and A100 nodes serve Gemma~3: 1B (three instances) and 4B (five) on the A30s, and 12B (four) and 27B (two, with TP2) on the A100s in BF16. Because vLLM supports Gemma~3 only in BF16 or FP32, V100 nodes are excluded due to lack of hardware BF16 support.

\begin{table}[t]
\centering\small
\caption{The Qwen2.5 cluster per tier: instances, the TPOT \sys uses (ms per token), prefill time (base in ms $+$ per prompt token in $\mu$s) and price (USD per million input/output tokens, from \citet{together} for the 3B and \citet{novita} for the rest).}
\label{tab:cluster}
\begin{adjustbox}{max width=\columnwidth}
\begin{tabular}{llrrrr}
\toprule
Model & GPU & Inst. & TPOT & Prefill & Price \\
\midrule
Qwen2.5-72B & A100$\times$4 & 2 & 44.3 & 78 + 131 & 0.38/0.40 \\
Qwen2.5-14B & V100$\times$4 & 3 & 20.9 & 33 + 129 & 0.15/0.15 \\
Qwen2.5-7B & A30 & 5 & 22.0 & 34 + 108 & 0.07/0.07 \\
Qwen2.5-3B & A30 & 3 & 11.4 & 18 + 54 & 0.06/0.06 \\
\bottomrule
\end{tabular}
\end{adjustbox}
\end{table}

\textbf{Data and quality.} Prompts are drawn from seven public benchmarks: RewardBench~\cite{lambert2024rewardbench}, CodeUltraFeedback~\cite{weyssow2024codeultrafeedback}, BeaverTails~\cite{ji2023beavertails}, MixInstruct~\cite{jiang2023llmblender}, LMSYS-Chat-1M~\cite{zheng2024lmsyschat1m}, GSM8K~\cite{cobbe2021trainingverifierssolvemath}, and SQuAD~\cite{rajpurkar2016squad}. We construct the evaluation and training pool by extending and modifying the dataset mixing pipeline from BEST-Route to incorporate the latter three domains. The training split (14{,}908\ prompts for Qwen, 8{,}086\ for Gemma) labels \sys's estimator and trains the routers. The
evaluation uses 3{,}532 held-out prompts for Qwen and 1{,}979 for Gemma. Every model's answer to every prompt is scored by DeepEval's G-Eval~\cite{deepeval} with Llama-3.1-8B-Instruct as the judge, against each prompt's reference answer, on a $[0,1]$ scale. For harmful prompts, the reference is a refusal.

\textbf{Load and runs.}  Requests arrive as a Poisson process at a fixed mean rate $\lambda$ of 12, 18, 24, 30, 36 or 42 req/s in the main comparison. Each run sends every evaluation prompt once, with the same arrivals and prompt order for every system. Each configuration runs once at each rate, and a subset also runs at two more seeds (\S\ref{sec:generality}) to assess variation across runs. Decoding is the same for every system: temperature 0, a frequency penalty of 1.2 and at most 1{,}024 output tokens. Before each run, the scheduler process restarts, every engine's prefix cache is reset, and the engines are drained and warmed up. Every engine also runs the same launch settings in all runs of a comparison.

\textbf{Metrics.} We report quality, mean E2E latency as the client sees it, including every wait in the scheduler, and cost per million requests at each tier's per-token prices (Table~\ref{tab:cluster}). A configuration \emph{beats} another when it is at least as good on all three and better on one. We also compare at equal quality. To compare systems at iso-quality, we construct \sys's Pareto frontier of quality against mean latency and cost across its presets at each arrival rate. For each baseline setting, we evaluate \sys's latency and cost at the baseline's quality via linear interpolation between the two adjacent frontier points. Baselines with quality below \sys's lowest preset are compared directly against that preset, while those exceeding its highest preset by more than 0.005 are omitted.

\textbf{Configurations.} A \emph{baseline setting} is one router setting on one dispatcher, or a standalone dispatcher at one rate as \S \ref{sec:impl}. AvengersPro runs at five settings and BEST-Route at four thresholds, each on LOR, RR, Po2 and LL, and the five dispatchers also run alone, which gives 41 settings at each rate (\S\ref{sec:impl}). The vLLM Semantic Router, evaluated on the same cluster with its shipped policy updated only with our quality scores, saturates at 3.3 req/s under rate of 6 (below our lowest evaluation rate) and is deferred to Appendix~\ref{app:vsr}. We show \sys at eight weight presets, written quality/latency/cost: uniform, 50/30/20, 60/20/20, 80/10/10 (the default), 90/5/5, 5/25/70, 22/18/60 and 25/5/70.

\subsection{Quality, latency and cost}\label{sec:main}

Across its eight presets, \sys beats 231 of the 246 baseline settings on quality, mean latency and cost together (Figure~\ref{fig:frontier}, Table~\ref{tab:main}). It beats every AvengersPro setting and every BEST-Route threshold above 0 on every dispatcher at every rate, and from 36 req/s it beats every baseline setting. At equal quality, \sys has the lower mean latency against 244 of the 246 baseline settings and the lower cost against 237 baselines. The median reduction is 37\% in latency and 32\% in cost, and the latency reduction grows with the load, from a median of 23\% at 12 req/s to 60\% at 42 req/s. At equal quality, its p99 latency is also lower by a median of 37\% (full results in Appendix~\ref{app:tail}).

\begin{table}[t]
\centering\small
\caption{\sys at its eight presets against the 41 baseline settings at each rate: the settings it beats, and its median and largest reductions of mean E2E latency and cost at equal quality.}
\label{tab:main}
\begin{adjustbox}{max width=\columnwidth}
\begin{tabular}{rrrrrr}
\toprule
$\lambda$ & Beaten & \multicolumn{2}{c}{E2E reduction} & \multicolumn{2}{c}{Cost reduction} \\ & & median & largest & median & largest \\
\midrule
12 & 37 of 41 & 23\% & 54\% & 26\% & 54\% \\
18 & 37 of 41 & 27\% & 82\% & 31\% & 54\% \\
24 & 36 of 41 & 31\% & 91\% & 32\% & 54\% \\
30 & 39 of 41 & 37\% & 92\% & 32\% & 54\% \\
36 & 41 of 41 & 47\% & 93\% & 37\% & 54\% \\
42 & 41 of 41 & 60\% & 93\% & 33\% & 53\% \\
\bottomrule
\end{tabular}
\end{adjustbox}
\end{table}

AvengersPro's latency grows with the load because its router chooses the model without reading the load. At 0.80, it sends
55\% of requests to the 72B model at every rate, and its mean latency rises from 5.92 s at 12 req/s to 14.27 s at 42 req/s (Figure~\ref{fig:frontier}, third row), while \sys at 80/10/10 stays
between 4.58 and 4.88 s at no lower quality. Also, \sys does not beat 15 settings (Appendix~\ref{app:open} lists them). They are BEST-Route at threshold~0 on every dispatcher at 12--24 req/s and on LL at 30 req/s, and the LL dispatcher alone at 24 and 30 req/s. At threshold 0, BEST-Route routes at least 90\% of requests to the 3B and 7B models, trading quality for cost savings within the coverage of \sys presets.

\subsection{The joint decision against its two variants}\label{sec:split}

\begin{figure*}[t]
\centering
\includegraphics[width=0.9\textwidth]{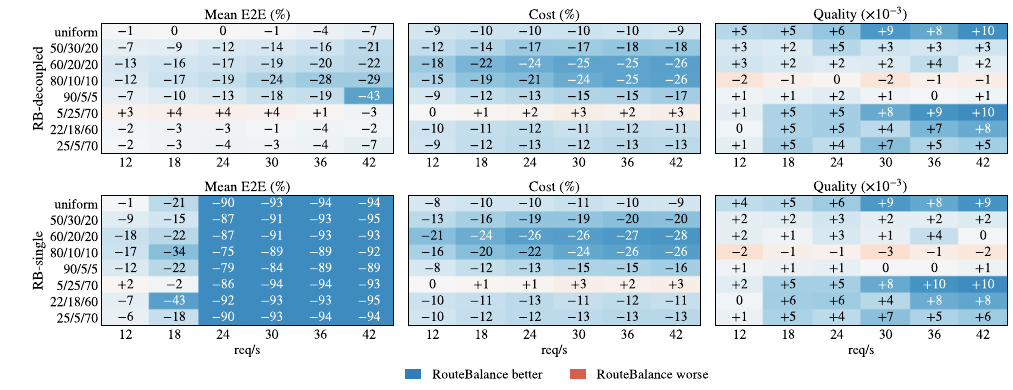}
\caption{\sys against RB-decoupled (top) and RB-single (bottom) at the same weights, at every preset and rate: \sys's change of mean E2E latency and cost (\%) and its difference in quality in thousandths. A '0' indicates a negligible change (under 0.5\% or a quality difference below 0.0005). Table~A5 gives every value.}
\label{fig:variants}
\Description{Heatmaps comparing RouteBalance with RB-decoupled and RB-single at the same weight presets and request rates.}
\end{figure*}

Figure~\ref{fig:variants} compares \sys with RB-decoupled and RB-single (\S\ref{sec:impl}) at every preset and rate. Against RB-decoupled, \sys is faster in 43 of the 48 preset--rate pairs by up to 43\%, and cheaper in 42 by up to 26\%, with quality differences from RB-decoupled between $-$0.0022 and +0.0101. At 80/10/10, \sys is 12--29\% faster and 15--26\% cheaper at 12, 30 and 42 req/s. The two also differ in where they send requests: RB-decoupled sends 48\% of requests to the 72B model at every rate, whereas \sys sends 39\% at 12 req/s and 32\% at 42 req/s, with quality within 0.0022.

RB-single falls further behind. \sys is faster in 47 of the 48 pairs by up to
95\%, and cheaper in 42 by up to 28\%, and RB-single beats it in none. At 80/10/10, \sys is 17--92\% faster at 12, 30 and 42 req/s and 16--26\% cheaper. RB-single also keeps 48\% of requests on the 72B model at every rate. From 24 req/s, its requests queue for their decisions. They wait 4.4 s or more at every seed, and up to 59 s at 42 req/s as missing overhead amortizing. Even net of that wait, RB-single is still 14--30\% slower at 80/10/10 at every rate and seed. At 5/25/70, RB-single is up to 2.9\% cheaper at up to 0.0102 lower quality, so neither beat the other.

\subsection{Sources of gains and scheduling overhead}\label{sec:ablation}

We also examine which design choices contribute to \sys's performance gains. Each variant below changes one component under the 80/10/10 preset, with detailed results reported in Appendices~\ref{app:ablation} and~\ref{app:reqnorm}.

The normalization over the round lets the load move requests between models. When each request is normalized over its own candidates instead, the eight presets beat 147 of the 246 settings instead of 231 (Appendix~\ref{app:reqnorm}), and \sys at 80/10/10 is up to 55\% slower, at 42 req/s. It then sends 48\% of requests to the 72B model at every rate, as RB-decoupled does, because the joint choice responds to the load through the round's range, which per-request normalization removes. At 80/10/10, the mean round grows from 1.8 requests at 12 req/s to 5.7 at 42 req/s, with quality range from 0.25 to 0.43. For requests where the 72B model has the highest estimated quality, its lead over the next-best model remains 0.076 on the judge's scale across all rates. As the arrival rate increases from 12 to 42 req/s, the round-wide quality range expands, reducing this advantage from 0.25 to 0.16 after normalization. Consequently, the fraction of these requests assigned to the 72B model decreases from 65\% to 53\% (Table~\ref{tab:roundrange}, Appendix~\ref{app:reqnorm}). The size and order of a round have smaller effects than its normalization. A fixed round of 32 requests or 200 ms changes mean latency by $-$4\% and $-$1\% and cost by $-$9\% and $-$5\% at 12 and 30 req/s, but latency by +30\% at 42 req/s. Taking a round's requests in arrival order instead of longest first changes it up to 2.5\%.

Without the latency term ($w_l{=}0$), \sys picks the model by quality and cost alone, and it is 12\% to 22\% slower and 14\% to 20\% more expensive. Replacing the analytical latency estimate with learned per-tier regressors trained on a load sweep increases mean latency by only 1\%. With sequential assignment and dead reckoning (\S\ref{sec:load}) in \sys, all requests except the first in each round are scored after accounting for earlier assignments, representing 74\% and 82\% of total decisions at 30 and 42 req/s, respectively. 6\% and 13\% of requests are assigned to instances already selected within the same round, showing the mitigation of request herding.

We also find that \sys's performance is far more sensitive to prompt-specific estimates than to the amount of labelled training data. Replacing per-prompt quality estimates with each model's mean label (prompt-blind estimates) increases \sys's latency by 2.2$\times$ to 15.8$\times$, because the estimator can no longer distinguish requests that smaller models can answer equally well. By contrast, just reducing the number of training labels to a quarter increases latency by at most 18.4\%. Furthermore, excluding either GSM8K or LMSYS training prompts to evaluate cross-domain robustness changes quality by at most 0.006 and latency by at most 5\%.

We analyze the overhead of scheduling as well. The time a request spends in \sys before it is dispatched grows from 123 ms at 12 req/s to 198 ms at 42 req/s (Table~\ref{tab:overhead}). It consists of the round's wait, the batched estimate and load read, both shared and amortized across the round's requests, and at most 1.2 ms of scheduling. Even with up to 32 concurrent routing calls, as in the main comparison, per-request routers incur increasing wait times under load: AvengersPro at 0.80 waits from 102 to 418 ms, while BEST-Route at 0.5 reaches 2{,}215 ms. The vLLM Semantic Router likewise processes requests independently, executing its classifier and embedding models to generate routing signals for each request on the same CPU, while allowing concurrent routing calls. At 6 req/s, this signal phase accounts for 99.0\% of a request's time in Envoy, and at 9 req/s and above, the router exceeds the default thread limit (detailed in Appendix~\ref{app:vsr}).

\subsection{Generality}\label{sec:generality}
\begin{table}[t]
\centering\small
\caption{Mean time a request spends in the scheduler before dispatch (ms): \sys at 80/10/10 by stage (the round's wait, batched estimate, load read and scheduling), compared to AvengersPro 0.80 (AP) and BEST-Route 0.5 (BR) both on LOR.}
\label{tab:overhead}
\begin{adjustbox}{max width=\columnwidth}
\begin{tabular}{rrrrrrrr}
\toprule
$\lambda$ & Wait & Estimate & Load & Schedule & Total & AP & BR \\
\midrule
12 & 53 & 47 & 22 & 0.6 & 123 & 102 & 165 \\
18 & 54 & 60 & 24 & 0.7 & 139 & 129 & 309 \\
24 & 55 & 69 & 24 & 0.6 & 149 & 154 & 697 \\
30 & 60 & 83 & 24 & 0.8 & 168 & 199 & 1{,}657 \\
36 & 69 & 96 & 28 & 1.2 & 194 & 305 & 2{,}215 \\
42 & 69 & 99 & 28 & 1.0 & 198 & 418 & 1{,}856 \\
\bottomrule
\end{tabular}
\end{adjustbox}
\end{table}

\begin{table*}[t]
\centering\footnotesize
\caption{Mean E2E latency (s) under gamma arrivals of shape $\alpha$: \sys, RB-decoupled (Dec.) and RB-single (Single) at 80/10/10, and AvengersPro 0.80 on LOR (AP).}
\label{tab:burst}
\begin{adjustbox}{max width=\textwidth}
\begin{tabular}{lrrrrrrrrrrrr}
\toprule
$\alpha$ & \multicolumn{4}{c}{12 req/s} & \multicolumn{4}{c}{30 req/s} & \multicolumn{4}{c}{42 req/s} \\ & \sys & Dec. & AP & Single & \sys & Dec. & AP & Single & \sys & Dec. & AP & Single \\
\midrule
0.25 & 4.45 & 5.31 & 5.93 & 5.57 & 4.51 & 6.22 & 7.51 & 41.39 & 4.78 & 6.95 & 13.99 & 54.90 \\
0.5 & 4.61 & 5.32 & 5.93 & 5.42 & 4.62 & 6.12 & 7.47 & 29.45 & 4.85 & 6.90 & 13.94 & 54.30 \\
1 (Poisson) & 4.69 & 5.31 & 5.92 & 5.66 & 4.66 & 6.12 & 7.33 & 40.77 & 4.88 & 6.82 & 14.27 & 63.99 \\
2 & 4.70 & 5.29 & 5.90 & 5.41 & 4.60 & 6.18 & 7.29 & 49.26 & 4.88 & 6.97 & 13.13 & 55.21 \\
\bottomrule
\end{tabular}
\end{adjustbox}
\end{table*}

\begin{table}[t]
\centering
\caption{Per-request budgets at 16 req/s: served quality (Q)
and percentage of requests exhausting their budgets (Exh.).}
\label{tab:budget}
\small
\setlength{\tabcolsep}{4pt}
\renewcommand{\arraystretch}{1.12}
\begin{adjustbox}{max width=\columnwidth}
\begin{tabular}{@{}lrrrrrr@{}}
\toprule
& \multicolumn{2}{c}{\sys{} + Filter}
& \multicolumn{2}{c}{\sys{} (No Filter)}
& \multicolumn{2}{c}{BEST-Route 0} \\
\cmidrule(lr){2-3}
\cmidrule(lr){4-5}
\cmidrule(l){6-7}
Budgeted & Q & Exh. & Q & Exh. & Q & Exh. \\
\midrule
75\% & \textbf{0.265} & \textbf{30.4}
     & 0.248 & 40.3 & 0.261 & 32.8 \\
45\% & \textbf{0.317} & \textbf{18.5}
     & 0.306 & 24.8 & 0.311 & 19.7 \\
30\% & \textbf{0.347} & \textbf{12.5}
     & 0.340 & 16.2 & 0.336 & 13.4 \\
\bottomrule
\end{tabular}
\end{adjustbox}
\end{table}

\begin{figure*}[t]
\centering
\includegraphics[width=0.85\textwidth]{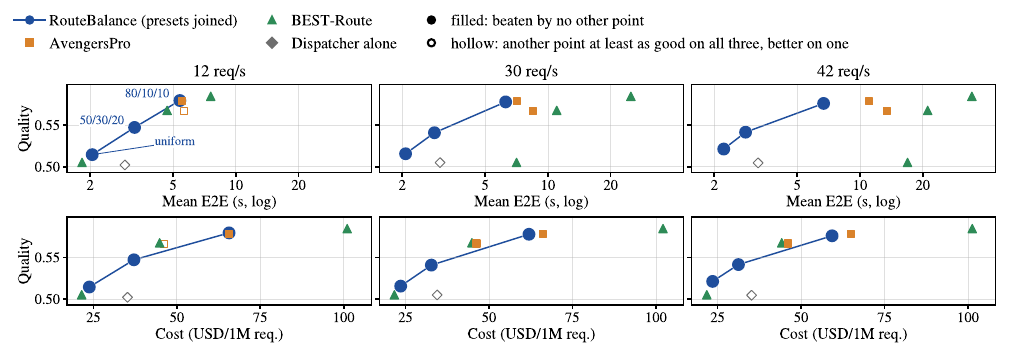}
\caption{Gemma~3 at 12, 30 and 42 req/s and OpenRouter prices~\cite{openrouter}: quality against mean latency or cost.}
\label{fig:gemma}
\Description{Gemma 3 quality plotted against mean latency and serving cost at 12, 30 and 42 requests per second.}
\end{figure*}

\textbf{A second model family.} We run the comparison on Gemma~3 with routing settings of Qwen. \sys runs at its default weights (80/10/10) and at the two balanced presets (50/30/20 and uniform). Each baseline runs at its balanced Qwen setting or the setting just below \sys's default quality on Qwen (AvengersPro 0.53 and 0.80, BEST-Route 0.3 and 0.5), and BEST-Route also runs at threshold~0. Both routers dispatch with LOR. No baseline setting beats any \sys preset on quality, latency and cost together at any rate (Figure~\ref{fig:gemma}). At equal quality, \sys reduces p99 E2E latency by a median of 24\% and 45\% across baseline settings at 30 and 42 req/s, respectively, while remaining within 1.5\% at 12 req/s (Appendices~\ref{app:gemma} and~\ref{app:tail}).

Against AvengersPro at 0.80, \sys at its default weights stays within 0.5\% of the quality. It is within 0.2\% of the cost at 12 req/s and 6\% and 9\% cheaper at 30 and 42 req/s, and its latency advantage grows with the load, from 2\% at 12 req/s to 40\% at 42 req/s. The cost margin is smaller than Qwen results because Gemma's 12B matches its 27B on 72\% of prompts, with mean quality 0.010 apart. So at high quality, every router picks between near-equivalent models. BEST-Route at 0.5 gains 0.9--1.5\% in quality by sending more requests to the 27B model, but at 1.4--5.1$\times$ \sys's latency and 1.5--1.7$\times$ its cost. Under load (30 and 42 req/s), each of the remaining baseline settings is slower than a \sys preset of at least its quality. Also on Gemma, \sys is faster and cheaper than both its variants at the same weights: up to 53\% faster than RB-decoupled and 89\% faster than RB-single, for at most 3.4\% and 4.4\% lower quality. 

\textbf{Bursty arrivals.} In Table~\ref{tab:burst}, gamma inter-arrival times of shape $\alpha$ replace the Poisson arrivals at the same mean rate for the Qwen experiment, the way vLLM's benchmark generates bursts (a smaller $\alpha$ is burstier). \sys at 80/10/10 stays at 4.4--5.0 s mean latency at every burstiness level, rates and seeds, against 5.2--7.9 s for RB-decoupled, 5.2--64.0 s for RB-single and 5.8--14.4 s for AvengersPro~0.80.

\textbf{Per-request budgets.} Table~\ref{tab:budget} evaluates the admission filter of \S\ref{sec:objective} at 16 req/s: 75\%, 45\%, or 30\% of requests carry a cost budget, and quality is judged on each request's served answer, with its output length capped by the budget. By filtering out candidate models whose expected costs exceed the budget, \sys achieves higher quality and a lower budget-exhaustion rate than BEST-Route at threshold~0. Compared with \sys without the filter, the admission filter reduces budget exhaustion from 40.3\% to 30.4\% when 75\% of requests are budgeted.

\textbf{Repeated runs.} We repeat all \sys presets and 85 selected Qwen baseline settings with two additional seeds, varying arrival times and prompt order. The baseline settings cover the seed-1 Pareto frontier, router knee points, and all standalone dispatchers. All Gemma settings are also repeated (Table~\ref{tab:spread}). Across seeds, the median / maximum relative range of mean E2E latency is 1.4\% / 6.8\% for \sys versus 2.7\% / 41.0\% for baselines on Qwen, and 2.3\% / 4.6\% versus 3.3\% / 19.4\% on Gemma. The maximum relative cost variation is also lower on Qwen (2.7\% vs. 11.6\%) and comparable on Gemma (3.6\% vs. 3.7\%). (see Appendix~\ref{app:seeds} for detailed results from seeds 2 and 3).

\begin{table}[t]
\centering\small
\caption{Spread across seeds (median / max): absolute quality range and relative ranges of E2E and cost as $(\max-\min)/\mathrm{mean}$.}
\label{tab:spread}
\begin{adjustbox}{max width=\columnwidth}
\begin{tabular}{lrccc}
\toprule
Settings & $n$ & Quality & E2E (\%) & Cost (\%) \\
\midrule
Qwen, \sys & 48 & 0.002 / 0.006 & 1.4 / 6.8 & 1.2 / 2.7 \\
Qwen, baselines & 85 & 0.000 / 0.011 & 2.7 / 41.0 & 1.0 / 11.6 \\
Gemma, \sys & 9 & 0.003 / 0.006 & 2.3 / 4.6 & 1.9 / 3.6 \\
Gemma, baselines & 18 & 0.000 / 0.008 & 3.3 / 19.4 & 1.1 / 3.7 \\
\bottomrule
\end{tabular}
\end{adjustbox}
\end{table}

\FloatBarrier

\section{Conclusion}\label{sec:conclusion}

\sys fuses model routing and load balancing into one decision per request through adaptive scheduling rounds, achieving lower latency and serving cost at equal quality than existing model routers and load balancers. This work opens up new opportunities for LLM serving systems that jointly optimize model routing, load balancing, and cost.

\bibliographystyle{ACM-Reference-Format}
\bibliography{references}

\appendix

\setcounter{table}{0}
\setcounter{figure}{0}
\renewcommand{\thetable}{A\arabic{table}}
\renewcommand{\thefigure}{A\arabic{figure}}
\renewcommand{\theHtable}{appendix.\arabic{table}}
\renewcommand{\theHfigure}{appendix.\arabic{figure}}
\section{Datasets}\label{app:data}

The prompts come from seven public datasets (Table~\ref{tab:datasets}), and each prompt carries its dataset's
reference answer. For the harmful prompts of BeaverTails, the reference is a refusal. The training and evaluation
prompts are disjoint. Every model's answer to every prompt is judged against the reference by DeepEval's G-Eval with
Llama-3.1-8B-Instruct, on a $[0,1]$ scale, and a prompt with an answer the judge could not score is dropped, which
leaves 3{,}532 evaluation prompts for Qwen.

\begin{table}[!htbp]\centering\small
\caption{Prompts per source dataset in each family's training and evaluation splits.}
\label{tab:datasets}
\begin{tabular}{lrrrr}
\toprule
 & \multicolumn{2}{c}{Qwen2.5} & \multicolumn{2}{c}{Gemma 3} \\
Dataset & Train & Eval & Train & Eval \\
\midrule
BeaverTails & 2{,}313 & 582 & 809 & 231 \\
CodeUltraFeedback & 2{,}242 & 518 & 246 & 75 \\
GSM8K & 2{,}361 & 495 & 1{,}841 & 408 \\
LMSYS-Chat-1M & 2{,}008 & 473 & 1{,}178 & 277 \\
MixInstruct & 2{,}300 & 567 & 1{,}119 & 280 \\
RewardBench & 1{,}336 & 325 & 549 & 127 \\
SQuAD & 2{,}348 & 572 & 2{,}344 & 581 \\
\midrule
Total & 14{,}908 & 3{,}532 & 8{,}086 & 1{,}979 \\
\bottomrule
\end{tabular}\end{table}

\section{Baselines and their settings}\label{app:baselines}

\paragraph{AvengersPro.} We run the authors' code with its default routing settings (the nearest
cluster, $\beta = 9$), on MiniLM embeddings of the prompts, and sweep its performance weight $pw$ over 0.25, 0.39, 0.53,
0.70 and 0.80. Its clusters and per-cluster scores are built from the same labelled training prompts as \sys's
estimator, with each model's cost at the fleet's prices.

\paragraph{BEST-Route.} We train the authors' DeBERTaV3-small ranker with their trainer on the same training prompts
and run their router with one sample per request at thresholds 0, 0.3, 0.5 and 0.7.

\paragraph{Prices.} Both routers use the fleet's per-token prices in their own cost terms. AvengersPro's cost term is
each model's average per-query cost within the cluster, as its authors compute it, and BEST-Route takes the cheapest
qualifying model by price.

\paragraph{Dispatchers.} LOR and Po2 read the engines' running and waiting counts from a cache refreshed every
50 ms. LOR gives a tie to the instance assigned least recently and then to the first in the fleet's order. Po2
compares two instances drawn at random. LL keeps the last 10 per-token latencies of each instance, as LiteLLM does,
and draws among tied instances at random.

\section{The routers one call at a time}\label{app:shipped}

The main comparison runs up to 32 router calls at once. Table~\ref{tab:shipped} also measures a serialized variant
of AvengersPro and BEST-Route, with one router call at a time. Each call makes a per-request model choice in both
variants; only the number of concurrent calls changes. On the same dispatcher, the serialized variant waits longer
at 30 and 42 req/s, and the eight presets beat 24 of its 27 settings at 12, 30 and 42 req/s.

\begin{table*}[!htbp]
\centering\footnotesize
\caption{The routers on LOR with up to 32 calls at once, as in the main comparison (32), and one call at a time (1): mean E2E latency and mean router wait, in seconds.}
\label{tab:shipped}
\begin{adjustbox}{max width=\textwidth}
\begin{tabular}{lrrrrrrrrrrrr}
\toprule
Router & \multicolumn{4}{c}{12 req/s} & \multicolumn{4}{c}{30 req/s} & \multicolumn{4}{c}{42 req/s} \\ & E2E 32 & E2E 1 & wait 32 & wait 1 & E2E 32 & E2E 1 & wait 32 & wait 1 & E2E 32 & E2E 1 & wait 32 & wait 1 \\
\midrule
AvengersPro 0.25 & 2.67 & 2.68 & 0.08 & 0.08 & 3.03 & 33.41 & 0.17 & 30.73 & 4.04 & 50.85 & 0.48 & 48.15 \\
AvengersPro 0.39 & 2.69 & 2.70 & 0.08 & 0.09 & 3.32 & 36.67 & 0.18 & 33.93 & 4.56 & 47.37 & 0.37 & 44.57 \\
AvengersPro 0.53 & 3.23 & 3.22 & 0.10 & 0.10 & 4.08 & 38.45 & 0.19 & 35.15 & 7.22 & 51.74 & 0.58 & 48.35 \\
AvengersPro 0.70 & 4.87 & 4.87 & 0.10 & 0.08 & 5.53 & 36.62 & 0.19 & 31.58 & 6.96 & 54.55 & 0.50 & 49.49 \\
AvengersPro 0.80 & 5.92 & 5.95 & 0.10 & 0.09 & 7.33 & 35.28 & 0.20 & 28.91 & 14.27 & 54.20 & 0.42 & 47.52 \\
BEST-Route 0 & 2.44 & 2.43 & 0.16 & 0.16 & 5.49 & 68.80 & 2.58 & 66.47 & 22.08 & 75.07 & 17.00 & 72.74 \\
BEST-Route 0.3 & 4.23 & 4.85 & 0.16 & 0.68 & 7.57 & 74.87 & 2.66 & 70.73 & 24.00 & 95.26 & 12.00 & 91.04 \\
BEST-Route 0.5 & 7.19 & 7.44 & 0.17 & 0.21 & 16.04 & 76.85 & 1.66 & 69.34 & 32.94 & 86.47 & 1.86 & 79.11 \\
BEST-Route 0.7 & 10.67 & 10.93 & 0.17 & 0.23 & 65.74 & 74.92 & 0.90 & 62.73 & 83.28 & 95.51 & 0.99 & 83.61 \\
\bottomrule
\end{tabular}
\end{adjustbox}
\end{table*}
{}

\section{The latency calibration}\label{app:calibration}

Each tier's TPOT is measured on the serving stack in a closed loop, on training prompts, after every engine's prefix
cache is reset. The levels run 1, 4, 16, 32, 64 and 96 requests at once on each instance, with 128 measured requests per
level. Every level is measured except 96 on the 72B model, whose KV cache does not fit 96 requests at once. \sys's
per-token decode time $\tau_t$ is the median TPOT of all the tier's samples over every level (Table~\ref{tab:calib}),
and the prefill line is fitted on the single-request measurements from the same runs.

\begin{table*}[!htbp]
\centering\footnotesize
\caption{The TPOT calibration: the median per-token decode time (ms) with $n$ requests running on each instance, the per-token decode time $\tau$ that \sys uses (the median of all the tier's samples over every level), the number of samples, and the $R^2$ of the prefill line.}
\label{tab:calib}
\begin{adjustbox}{max width=\textwidth}
\begin{tabular}{lrrrrrrrrr}
\toprule
Model & $n{=}1$ & $n{=}4$ & $n{=}16$ & $n{=}32$ & $n{=}64$ & $n{=}96$ & $\tau$ & Samples & Prefill $R^2$ \\
\midrule
Qwen2.5-72B & 43.6 & 44.6 & 44.4 & 43.4 & 50.6 & -- & 44.3 & 1{,}280 & 0.86 \\
Qwen2.5-14B & 11.8 & 15.2 & 18.9 & 23.6 & 34.9 & 46.8 & 20.9 & 2{,}304 & 0.79 \\
Qwen2.5-7B & 18.5 & 19.3 & 21.1 & 22.9 & 28.1 & 36.2 & 22.0 & 3{,}840 & 0.93 \\
Qwen2.5-3B & 9.6 & 9.8 & 10.8 & 12.0 & 15.3 & 19.7 & 11.4 & 2{,}304 & 0.86 \\
\bottomrule
\end{tabular}
\end{adjustbox}
\end{table*}
{}

\section{Per-request budgets}\label{app:budget}

A request may carry a cost budget. \sys then admits a candidate instance only if the expected cost fits within it.
The expected cost is the request's prompt at the tier's input price plus the model's expected output length, capped at the request's maximum
tokens, at the tier's output price. When no candidate fits, the candidates with the lowest worst-case cost (the prompt
plus the maximum tokens) remain. Every system caps each request's output at what its budget can pay for on the chosen
tier. The lowest of the four budget levels admits only the 3B model, and the other three admit, in turn, the 7B, 14B
and 72B models as well. In the runs, 75\%, 45\% or 30\% of the requests carry a budget. Quality is judged on the served
text: an empty reply, or one of the 5 served texts the judge could not score, scores 0. A request
exhausts its budget when its answer is cut at the budget or the budget cannot pay for its prompt.

\section{Settings no preset beats}\label{app:open}

Table~\ref{tab:open} lists every baseline setting of the main comparison that no \sys preset beats on quality, latency and
cost together, beside the \sys preset nearest to it in quality.

\begin{table*}[!htbp]
\centering\footnotesize
\caption{Baseline settings that no \sys preset beats on quality, latency and cost together, beside the \sys preset nearest in quality. Latency is the mean E2E in seconds and cost is USD per million requests.}
\label{tab:open}
\begin{adjustbox}{max width=\textwidth}
\begin{tabular}{rlrrrlrrr}
\toprule
$\lambda$ & Baseline & Q & E2E & Cost & \sys & Q & E2E & Cost \\
\midrule
12 & BEST-Route 0 / LOR & 0.3924 & 2.44 & 20.9 & 25/5/70 & 0.3885 & 2.53 & 21.6 \\
12 & BEST-Route 0 / RR & 0.3924 & 2.45 & 21.0 & 25/5/70 & 0.3885 & 2.53 & 21.6 \\
12 & BEST-Route 0 / Po2 & 0.3924 & 2.42 & 20.8 & 25/5/70 & 0.3885 & 2.53 & 21.6 \\
12 & BEST-Route 0 / LL & 0.3924 & 2.47 & 21.1 & 25/5/70 & 0.3885 & 2.53 & 21.6 \\
18 & BEST-Route 0 / LOR & 0.3924 & 2.66 & 21.0 & 25/5/70 & 0.3924 & 2.60 & 21.2 \\
18 & BEST-Route 0 / RR & 0.3924 & 2.67 & 20.8 & 25/5/70 & 0.3924 & 2.60 & 21.2 \\
18 & BEST-Route 0 / Po2 & 0.3924 & 2.66 & 20.9 & 25/5/70 & 0.3924 & 2.60 & 21.2 \\
18 & BEST-Route 0 / LL & 0.3924 & 2.65 & 21.0 & 25/5/70 & 0.3924 & 2.60 & 21.2 \\
24 & BEST-Route 0 / LOR & 0.3924 & 3.29 & 21.0 & 22/18/60 & 0.3918 & 2.49 & 20.6 \\
24 & BEST-Route 0 / RR & 0.3924 & 3.24 & 20.9 & 22/18/60 & 0.3918 & 2.49 & 20.6 \\
24 & BEST-Route 0 / Po2 & 0.3924 & 3.25 & 20.9 & 22/18/60 & 0.3918 & 2.49 & 20.6 \\
24 & BEST-Route 0 / LL & 0.3924 & 3.36 & 21.0 & 22/18/60 & 0.3918 & 2.49 & 20.6 \\
24 & LL alone & 0.3505 & 2.43 & 19.0 & 5/25/70 & 0.3704 & 2.46 & 18.8 \\
30 & BEST-Route 0 / LL & 0.3924 & 5.08 & 20.8 & 25/5/70 & 0.3940 & 2.73 & 21.0 \\
30 & LL alone & 0.3522 & 2.54 & 19.4 & 5/25/70 & 0.3730 & 2.57 & 19.0 \\
\bottomrule
\end{tabular}
\end{adjustbox}
\end{table*}
{}

\section{RB-decoupled and RB-single}\label{app:variants}

Table~\ref{tab:decoupled} gives every value that the main text's Figure~\ref{fig:variants} draws: \sys against each variant at the same
weights, at every preset and rate.

\begin{table*}[!htbp]
\centering\footnotesize\setlength{\tabcolsep}{3.5pt}
\caption{\sys against RB-decoupled and RB-single at the same weights, at every preset and rate: \sys's change of mean E2E latency and cost (\%) and its difference in quality, in thousandths, as in the main text's Figure~\ref{fig:variants}.}
\label{tab:decoupled}
\begin{adjustbox}{max width=\textwidth}
\begin{tabular}{lrrrrrrrrrrrrrrrrrr}
\toprule
Preset & \multicolumn{3}{c}{12 req/s} & \multicolumn{3}{c}{18 req/s} & \multicolumn{3}{c}{24 req/s} & \multicolumn{3}{c}{30 req/s} & \multicolumn{3}{c}{36 req/s} & \multicolumn{3}{c}{42 req/s} \\ \cmidrule(lr){2-4}\cmidrule(lr){5-7}\cmidrule(lr){8-10}\cmidrule(lr){11-13}\cmidrule(lr){14-16}\cmidrule(lr){17-19} & E2E & Cost & Q & E2E & Cost & Q & E2E & Cost & Q & E2E & Cost & Q & E2E & Cost & Q & E2E & Cost & Q \\
\midrule
\multicolumn{19}{l}{\emph{RB-decoupled}} \\
uniform & $-$1 & $-$9 & +5 & 0 & $-$10 & +5 & 0 & $-$10 & +6 & $-$1 & $-$10 & +9 & $-$4 & $-$10 & +8 & $-$7 & $-$9 & +10 \\
50/30/20 & $-$7 & $-$12 & +3 & $-$9 & $-$14 & +2 & $-$12 & $-$17 & +5 & $-$14 & $-$17 & +3 & $-$16 & $-$18 & +3 & $-$21 & $-$18 & +3 \\
60/20/20 & $-$13 & $-$18 & +3 & $-$16 & $-$22 & +2 & $-$17 & $-$24 & +2 & $-$19 & $-$25 & +2 & $-$20 & $-$25 & +4 & $-$22 & $-$26 & +2 \\
80/10/10 & $-$12 & $-$15 & $-$2 & $-$17 & $-$19 & $-$1 & $-$19 & $-$21 & 0 & $-$24 & $-$24 & $-$2 & $-$28 & $-$25 & $-$1 & $-$29 & $-$26 & $-$1 \\
90/5/5 & $-$7 & $-$9 & +1 & $-$10 & $-$12 & +1 & $-$13 & $-$13 & +2 & $-$18 & $-$15 & +1 & $-$19 & $-$15 & 0 & $-$43 & $-$17 & +1 \\
5/25/70 & +3 & 0 & +1 & +4 & +1 & +5 & +4 & +2 & +5 & +4 & +3 & +8 & +1 & +2 & +9 & $-$3 & +3 & +10 \\
22/18/60 & $-$2 & $-$10 & 0 & $-$3 & $-$11 & +5 & $-$3 & $-$12 & +5 & $-$1 & $-$11 & +4 & $-$4 & $-$12 & +7 & $-$2 & $-$11 & +8 \\
25/5/70 & $-$2 & $-$9 & +1 & $-$3 & $-$12 & +5 & $-$4 & $-$13 & +4 & $-$3 & $-$12 & +7 & $-$4 & $-$13 & +5 & $-$7 & $-$13 & +5 \\
\multicolumn{19}{l}{\emph{RB-single}} \\
uniform & $-$1 & $-$8 & +4 & $-$21 & $-$10 & +5 & $-$90 & $-$10 & +6 & $-$93 & $-$11 & +9 & $-$94 & $-$10 & +8 & $-$94 & $-$9 & +9 \\
50/30/20 & $-$9 & $-$13 & +2 & $-$15 & $-$16 & +2 & $-$87 & $-$19 & +3 & $-$91 & $-$19 & +2 & $-$93 & $-$20 & +2 & $-$95 & $-$20 & +2 \\
60/20/20 & $-$18 & $-$21 & +2 & $-$22 & $-$24 & +1 & $-$87 & $-$26 & +3 & $-$91 & $-$26 & +1 & $-$93 & $-$27 & +4 & $-$93 & $-$28 & 0 \\
80/10/10 & $-$17 & $-$16 & $-$2 & $-$34 & $-$20 & $-$1 & $-$75 & $-$22 & $-$1 & $-$89 & $-$24 & $-$3 & $-$89 & $-$26 & $-$1 & $-$92 & $-$26 & $-$2 \\
90/5/5 & $-$12 & $-$8 & +1 & $-$22 & $-$12 & +1 & $-$79 & $-$13 & +1 & $-$84 & $-$15 & 0 & $-$89 & $-$15 & 0 & $-$89 & $-$16 & +1 \\
5/25/70 & +2 & 0 & +2 & $-$2 & +1 & +5 & $-$86 & +1 & +5 & $-$94 & +3 & +8 & $-$94 & +2 & +10 & $-$93 & +3 & +10 \\
22/18/60 & $-$7 & $-$10 & 0 & $-$43 & $-$11 & +6 & $-$92 & $-$13 & +6 & $-$93 & $-$11 & +4 & $-$93 & $-$12 & +8 & $-$95 & $-$11 & +8 \\
25/5/70 & $-$6 & $-$10 & +1 & $-$18 & $-$12 & +5 & $-$90 & $-$12 & +4 & $-$93 & $-$13 & +7 & $-$94 & $-$13 & +5 & $-$94 & $-$13 & +6 \\
\bottomrule
\end{tabular}
\end{adjustbox}
\end{table*}
{}

\section{Where the gain comes from}\label{app:ablation}

Each variant changes one part of \sys at 80/10/10, and Table~\ref{tab:ablation} compares it with a reference run of
\sys at 80/10/10, at 12, 30 and 42 req/s.

\begin{table*}[!htbp]
\centering\footnotesize
\caption{Variants of \sys at 80/10/10 against a reference run of \sys at 80/10/10: the difference in quality, and the change (\%) of mean E2E latency and of cost ($^{*}$against \sys at 80/10/10 in the main comparison).}
\label{tab:ablation}
\begin{adjustbox}{max width=\textwidth}
\begin{tabular}{lrrrrrrrrr}
\toprule
Variant & \multicolumn{3}{c}{12 req/s} & \multicolumn{3}{c}{30 req/s} & \multicolumn{3}{c}{42 req/s} \\ & Q & E2E & Cost & Q & E2E & Cost & Q & E2E & Cost \\
\midrule
LPT off (arrival order) & 0.000 & 0 & 0 & 0.000 & $-$1 & $-$1 & +0.001 & $-$2 & $-$2 \\
Fixed round (32 requests or 200 ms) & 0.000 & $-$4 & $-$9 & $-$0.001 & $-$1 & $-$5 & +0.001 & +30 & 0 \\
Scores normalized per request$^{*}$ & +0.002 & +13 & +17 & +0.002 & +36 & +34 & +0.001 & +55 & +34 \\
No latency term ($w_l{=}0$) & +0.001 & +12 & +14 & +0.003 & +19 & +18 & +0.004 & +22 & +20 \\
Learned per-tier regressors & 0.000 & +1 & 0 & 0.000 & +1 & $-$2 & +0.001 & +1 & $-$1 \\
Prompt-blind estimates & +0.001 & +120 & +109 & +0.002 & +1{,}226 & +129 & +0.003 & +1{,}483 & +136 \\
75\% of the labels & $-$0.003 & +6 & +5 & $-$0.003 & +5 & +2 & $-$0.005 & +2 & +3 \\
50\% of the labels & +0.003 & +10 & +10 & 0.000 & +6 & +6 & +0.001 & +8 & +9 \\
25\% of the labels & +0.001 & +13 & +14 & 0.000 & +14 & +13 & 0.000 & +18 & +14 \\
Without GSM8K's prompts & $-$0.006 & +2 & +1 & $-$0.006 & +1 & $-$2 & $-$0.005 & $-$2 & $-$2 \\
Without LMSYS's prompts & $-$0.001 & +3 & +3 & 0.000 & +3 & +2 & $-$0.001 & +5 & +3 \\
\bottomrule
\end{tabular}
\end{adjustbox}
\end{table*}
{}

\section{Normalizing over the round}\label{app:reqnorm}

With each request's scores normalized over its own candidates instead of over the round, the eight presets beat
147 of the 246 baseline settings instead of 231. Table~\ref{tab:reqnorm}
gives the latency and the 72B model's share at three rates, beside RB-decoupled's latency.

\begin{table*}[!htbp]
\centering\footnotesize\setlength{\tabcolsep}{4pt}
\caption{\sys with its scores normalized over the round and over each request's own candidates, and RB-decoupled (Dec.): mean E2E latency (s) and the 72B model's share of requests (\%).}
\label{tab:reqnorm}
\begin{adjustbox}{max width=\textwidth}
\begin{tabular}{lrrrrrrrrrrrrrrr}
\toprule
Preset & \multicolumn{5}{c}{12 req/s} & \multicolumn{5}{c}{30 req/s} & \multicolumn{5}{c}{42 req/s} \\ \cmidrule(lr){2-6}\cmidrule(lr){7-11}\cmidrule(lr){12-16} & \multicolumn{3}{c}{E2E} & \multicolumn{2}{c}{72B} & \multicolumn{3}{c}{E2E} & \multicolumn{2}{c}{72B} & \multicolumn{3}{c}{E2E} & \multicolumn{2}{c}{72B} \\ \cmidrule(lr){2-4}\cmidrule(lr){5-6}\cmidrule(lr){7-9}\cmidrule(lr){10-11}\cmidrule(lr){12-14}\cmidrule(lr){15-16} & round & request & Dec. & round & request & round & request & Dec. & round & request & round & request & Dec. & round & request \\
\midrule
uniform & 2.49 & 2.57 & 2.51 & 5 & 1 & 2.74 & 2.87 & 2.78 & 6 & 1 & 2.92 & 3.19 & 3.14 & 7 & 1 \\
50/30/20 & 2.99 & 3.33 & 3.22 & 12 & 13 & 3.13 & 3.77 & 3.62 & 12 & 14 & 3.28 & 4.02 & 4.15 & 13 & 14 \\
60/20/20 & 3.36 & 4.03 & 3.89 & 18 & 25 & 3.44 & 4.36 & 4.23 & 17 & 25 & 3.57 & 4.89 & 4.60 & 16 & 26 \\
80/10/10 & 4.69 & 5.31 & 5.31 & 39 & 48 & 4.66 & 6.36 & 6.12 & 32 & 48 & 4.88 & 7.57 & 6.82 & 32 & 48 \\
90/5/5 & 5.40 & 5.83 & 5.79 & 48 & 55 & 5.82 & 7.15 & 7.06 & 45 & 55 & 6.43 & 11.72 & 11.25 & 44 & 54 \\
5/25/70 & 2.25 & 2.25 & 2.18 & 0 & 0 & 2.57 & 2.55 & 2.48 & 0 & 0 & 2.80 & 2.96 & 2.90 & 0 & 0 \\
22/18/60 & 2.35 & 2.52 & 2.41 & 1 & 0 & 2.62 & 2.73 & 2.65 & 2 & 0 & 2.81 & 2.87 & 2.88 & 2 & 0 \\
25/5/70 & 2.53 & 2.67 & 2.58 & 1 & 0 & 2.73 & 2.91 & 2.82 & 3 & 0 & 2.90 & 3.02 & 3.12 & 3 & 0 \\
\bottomrule
\end{tabular}
\end{adjustbox}
\end{table*}
{}

Table~\ref{tab:roundrange} follows the range itself at 80/10/10. The mean round grows with the rate, and the range of
estimated quality over every request of the round and every instance of the fleet widens with it. The 72B model is
estimated to answer 60\% of the requests best, and its lead over the next model on them is 0.076
on the judge's scale at every rate, because the estimates do not depend on the round, but as a share of each request's own
round's range it shrinks, and so does the share of those requests the 72B receives.

\begin{table}[!htbp]
\centering\small
\caption{The round's range at 80/10/10 (seed 1), from the decision logs: the mean round size, the mean quality range of the round, the 72B model's lead over the next model on the 60\% of requests it is estimated to answer best, on the judge's scale and as a share of the range of each request's own round, both averaged over those requests, the share of those requests it received (Kept), and its share of all requests.}
\label{tab:roundrange}
\begin{adjustbox}{max width=\columnwidth}
\begin{tabular}{rrrrrrr}
\toprule
$\lambda$ & Round & Range & \multicolumn{2}{c}{72B lead} & Kept & 72B \\ \cmidrule(lr){4-5} & size & & judge & of range & \% & \% \\
\midrule
12 & 1.8 & 0.25 & 0.076 & 0.25 & 65 & 39 \\
18 & 2.4 & 0.30 & 0.076 & 0.21 & 59 & 35 \\
24 & 3.1 & 0.34 & 0.076 & 0.19 & 57 & 34 \\
30 & 3.9 & 0.38 & 0.076 & 0.18 & 54 & 32 \\
36 & 5.0 & 0.42 & 0.076 & 0.17 & 53 & 32 \\
42 & 5.7 & 0.43 & 0.076 & 0.16 & 53 & 32 \\
\bottomrule
\end{tabular}
\end{adjustbox}
\end{table}
{}

\section{Gemma 3}\label{app:gemma}

Table~\ref{tab:gemma} gives every setting of the comparison on Gemma~3, with RB-decoupled and RB-single at the same weights. At
30 and 42 req/s, each baseline setting other than AvengersPro at 0.80 and BEST-Route at 0.5 is slower than a \sys preset of at least its quality: at 42 req/s, AvengersPro
0.53 and BEST-Route 0.3 take 2.0--3.2$\times$ the latency of \sys at
80/10/10, and BEST-Route~0 takes 7.6$\times$ that of \sys at uniform. Against the two
variants at the same weights, \sys is 10--53\% faster and
10--34\% cheaper than RB-decoupled, for
0.1--3.4\% lower quality, and 11--89\%
faster and 11--36\% cheaper than RB-single, for
0.1--4.4\% lower quality.

\begin{table*}[!htbp]
\centering\footnotesize
\caption{Gemma~3 with routing settings fixed on Qwen: quality, mean E2E latency (s) and cost (USD per million requests) at OpenRouter's token prices (\url{https://openrouter.ai/models}). OpenRouter does not list the 1B, which we price at \$0.05/\$0.05 per million input/output tokens, the lowest input price it lists for a Gemma 3 model.}
\label{tab:gemma}
\begin{adjustbox}{max width=\textwidth}
\begin{tabular}{lrrrrrrrrr}
\toprule
Setting & \multicolumn{3}{c}{12 req/s} & \multicolumn{3}{c}{30 req/s} & \multicolumn{3}{c}{42 req/s} \\ & Q & E2E & Cost & Q & E2E & Cost & Q & E2E & Cost \\
\midrule
\sys 80/10/10 & 0.579 & 5.39 & 65.6 & 0.578 & 6.27 & 62.0 & 0.576 & 6.69 & 59.3 \\
\sys 50/30/20 & 0.547 & 3.27 & 37.2 & 0.541 & 2.86 & 32.8 & 0.541 & 2.83 & 31.3 \\
\sys uniform & 0.514 & 2.05 & 23.8 & 0.516 & 2.08 & 23.6 & 0.521 & 2.22 & 23.5 \\
RB-decoupled 80/10/10 & 0.580 & 5.98 & 73.5 & 0.580 & 7.39 & 74.2 & 0.580 & 8.63 & 73.7 \\
RB-decoupled 50/30/20 & 0.560 & 4.40 & 45.8 & 0.560 & 5.26 & 46.6 & 0.560 & 6.02 & 47.0 \\
RB-decoupled uniform & 0.523 & 2.48 & 26.5 & 0.523 & 2.66 & 26.4 & 0.523 & 2.93 & 26.4 \\
AvengersPro 0.80 & 0.578 & 5.51 & 65.6 & 0.578 & 7.11 & 66.1 & 0.578 & 11.08 & 64.9 \\
AvengersPro 0.53 & 0.567 & 5.64 & 46.2 & 0.567 & 8.46 & 46.3 & 0.567 & 13.45 & 46.0 \\
BEST-Route 0.5 & 0.584 & 7.59 & 101.0 & 0.584 & 25.02 & 102.1 & 0.584 & 34.42 & 101.3 \\
BEST-Route 0.3 & 0.567 & 4.69 & 44.9 & 0.567 & 11.03 & 44.9 & 0.567 & 21.13 & 44.2 \\
BEST-Route 0 & 0.505 & 1.83 & 21.5 & 0.505 & 7.07 & 21.7 & 0.505 & 16.91 & 21.8 \\
LOR alone & 0.502 & 2.94 & 35.2 & 0.505 & 3.04 & 34.5 & 0.505 & 3.24 & 35.2 \\
RB-single 80/10/10 & 0.580 & 6.06 & 74.5 & 0.579 & 19.66 & 74.7 & 0.579 & 25.81 & 73.9 \\
RB-single 50/30/20 & 0.563 & 4.51 & 47.3 & 0.565 & 15.73 & 49.4 & 0.566 & 24.61 & 49.1 \\
RB-single uniform & 0.529 & 2.59 & 26.9 & 0.530 & 10.26 & 27.3 & 0.529 & 19.94 & 27.3 \\
\bottomrule
\end{tabular}
\end{adjustbox}
\end{table*}
{}

\begin{table*}[!htbp]
\centering\footnotesize
\caption{The vLLM Semantic Router, with its shipped policy changed only in its quality scores, beside \sys and round robin on the same cluster: requests completed, quality of the completed requests, mean E2E latency (s) and cost (USD per million requests).}
\label{tab:vsr}
\begin{adjustbox}{max width=\textwidth}
\begin{tabular}{lrrrrrrrr}
\toprule
System & \multicolumn{4}{c}{6 req/s} & \multicolumn{4}{c}{9 req/s} \\ & Done & Q & E2E & Cost & Done & Q & E2E & Cost \\
\midrule
\sys 80/10/10 & 3{,}532/3{,}532 & 0.432 & 5.01 & 61.5 & 3{,}532/3{,}532 & 0.431 & 4.93 & 59.6 \\
RR alone & 3{,}532/3{,}532 & 0.374 & 4.24 & 41.2 & 3{,}532/3{,}532 & 0.374 & 4.33 & 41.5 \\
vLLM-SR & 3{,}532/3{,}532 & 0.402 & 483.97 & 49.0 & 16/3{,}532 & 0.344 & 8.19 & 43.2 \\
\bottomrule
\end{tabular}
\end{adjustbox}
\end{table*}

\section{The vLLM Semantic Router}\label{app:vsr}

We deploy release v0.4.0 (commit \texttt{6573123}) in its maintained single-host setup, with its router and Envoy
as its command-line tool starts them, in front of the same engines. It routes with its built-in Mixture-of-Models
Balance policy, which is aware of quality, latency and cost. The policy first assigns each request to one of three
model pools (simple, medium, reasoning) from domain and fact-check classifiers, the prompt's embedding similarity to
example tasks, and keyword and structure rules. It then ranks the pool's models lexicographically. First comes each
model's quality score, keeping the models within 12\%, 7\% or 3\% of the best. Then come, in a pool-specific order,
the 95th-percentile first-token latency (per-token latency in the reasoning pool) that the router has observed for
each model and the request's estimated token cost. Last come the model's in-flight requests at the router.

We run the policy in one configuration, with its classifiers on the scheduler node's CPU, as every router here runs.
We assign the models to the pools by size, as neighbouring pairs (simple 3B and 7B, medium 7B and 14B, reasoning 14B
and 72B), give it the fleet's prices, and raise its request-path timeouts to the client's six hours, as for every
system. The shipped recipe takes the quality score from published benchmarks (MMLU-Pro in the simple and medium pools,
GPQA Diamond and Humanity's Last Exam in the reasoning pool). We supply instead each model's mean judged score on the
training split, the labels every other system here trains on, through the release's documented custom evaluations.
This is our only change to the shipped policy.

Table~\ref{tab:vsr} compares the vLLM Semantic Router (vLLM-SR) with \sys and round robin (RR) at 6 and
9 req/s. The router serves 3.3 requests per second when 6 arrive per second, with a mean latency of 484 s there, and at 9 req/s it stopped during
the run with 16 of 3{,}532 requests completed. At 6 req/s, \sys at 80/10/10 serves a
quality of 0.432 at a mean latency of 5.01 s, and round robin a quality of 0.374
at 4.24 s.

Like BEST-Route, the vLLM Semantic Router decides each request on its own, on the request path: every request runs its classifiers once,
on the scheduler node's CPU, whereas \sys, on the same CPU, runs its estimates once per round. Its requests therefore
wait in the router's signal phase, where its domain and fact-check classifiers and its embedding-similarity signals
run. At 6 req/s, a request spends 483.97 s in Envoy on average, 479.33 s of it
(99.0\%) in the signal phase and 3.5 ms (at most
65 ms) in the rest of its routing, the model choice included. At 9 to 30 req/s, the router
stopped when it exceeded Go's default limit of 10{,}000 threads, 84 to
328 s into the run. It was then holding 2{,}505 to 2{,}953 requests,
with 9{,}632 to 9{,}696 of its threads in native calls of its classifier and
embedding models, and it had completed 16 requests at 9 req/s, 4 at 12 and none at 18 to 30. No thread was in the model choice, the one step that reads the quality scores we supplied. On the same CPU, a
request spends 123--198 ms in \sys before it is dispatched, at 12 to 42 req/s.

\begin{table}[t]
\centering\small
\caption{p99 E2E latency: \sys's change at equal quality against the baseline settings (the median change and the largest reduction, \%, negative where \sys is lower) and p99 (s) at 80/10/10 for \sys, RB-decoupled (Dec.) and AvengersPro 0.80 on LOR (AP).}
\label{tab:tail}
\begin{adjustbox}{width=0.45\textwidth}
\begin{tabular*}{0.75\textwidth}{@{\extracolsep{\fill}}lrrrrr@{}}
\toprule
$\lambda$ & Median & Largest & \sys & Dec. & AP \\
\midrule
\multicolumn{6}{l}{\emph{Qwen2.5}} \\
12 & $-$27 & $-$59 & 26.8 & 34.1 & 34.5 \\
18 & $-$31 & $-$64 & 26.8 & 31.8 & 38.3 \\
24 & $-$32 & $-$74 & 29.2 & 35.2 & 44.5 \\
30 & $-$35 & $-$80 & 28.3 & 38.1 & 42.1 \\
36 & $-$50 & $-$83 & 26.7 & 40.7 & 49.1 \\
42 & $-$56 & $-$83 & 27.9 & 41.4 & 63.0 \\
\multicolumn{6}{l}{\emph{Gemma 3}} \\
12 & +1 & $-$41 & 25.5 & 30.7 & 24.6 \\
30 & $-$24 & $-$45 & 30.7 & 37.2 & 30.7 \\
42 & $-$45 & $-$59 & 32.0 & 40.6 & 39.5 \\
\bottomrule
\end{tabular*}
\end{adjustbox}
\end{table}
{}

\begin{figure*}[t]
\centering
\includegraphics[width=\textwidth]{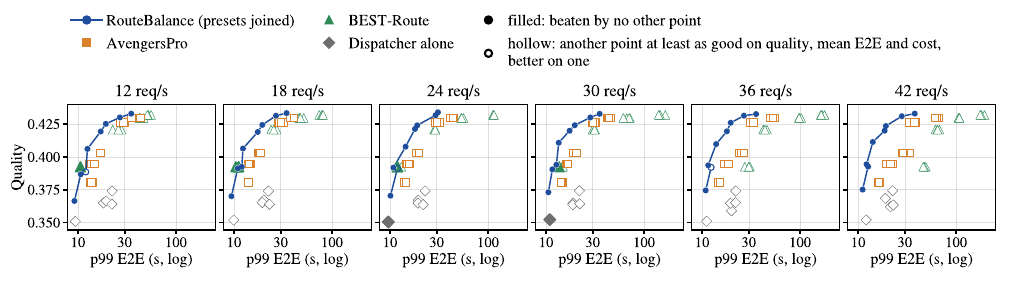}\\[2pt]
\includegraphics[width=\textwidth]{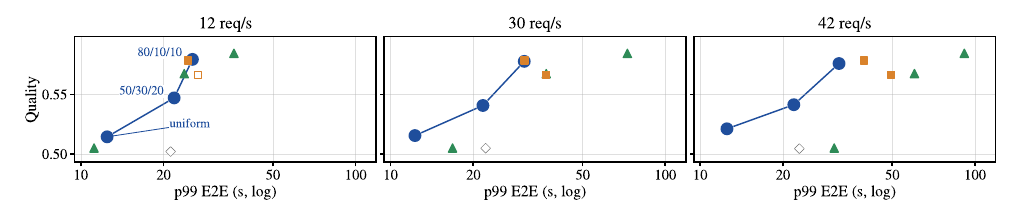}
\caption{Quality against p99 E2E latency (log scale): Qwen2.5 at each rate (top) and Gemma~3 at 12, 30 and 42 req/s
(bottom). The line joins \sys's presets, eight on Qwen and three on Gemma (weights labelled at 12 req/s). Filled and
hollow are decided on mean latency, as in the main text. On Qwen, a router's settings on its four dispatchers, and some
of \sys's presets, nearly coincide, and at 18 and 24 req/s one of \sys's presets coincides with BEST-Route at
threshold~0. On Gemma at 30 req/s, AvengersPro 0.80 lies inside \sys's 80/10/10 circle, both filled, and BEST-Route
0.3 is hidden under AvengersPro 0.53.}
\label{fig:tail}
\Description{Plots of quality against p99 end-to-end latency for Qwen2.5 and Gemma 3 across the evaluated request rates.}
\end{figure*}

\section{Repeated runs}\label{app:seeds}

\looseness=-1 This section gives each seed of the repeated Poisson runs on Qwen and Gemma and of the Qwen variant comparisons. At seeds 2 and 3, we rerun every
setting that no other setting beats, each system's setting closest to its best quality, latency and cost, every \sys
preset at each rate, RB-decoupled and RB-single at every preset and rate, and every Gemma setting. At seeds 1, 2 and 3, the presets beat
76, 65 and 74 of the
85 repeated Qwen baseline settings, and 3 of
18 on Gemma at every seed. Some settings are beaten at some seeds and not at others. At each seed where such a setting is not beaten, a
preset that is no worse on the other two measures is within 0.2\% of it in quality,
1.4\% in latency or 0.7\% in cost. Against the repeated baseline
settings, the median reduction of \sys's latency at equal quality is 34\%,
35\% and 35\% at seeds 1--3 on Qwen and 32\%, 32\% and 31\% on Gemma.
At seeds 2 and 3, \sys is faster than RB-decoupled in 43 and 43 of the
48 pairs and than RB-single in 48 and 47, and it is cheaper in
42 and 42 pairs against RB-decoupled and 42 and
42 against RB-single.

The baselines' largest ranges of mean latency across the seeds,
41.0\% for BEST-Route 0 / LL on Qwen and 19.4\% for
BEST-Route 0 / LOR on Gemma, both at 30 req/s, come from the router wait: of their
2.31 s and 1.32 s ranges, 2.27 s and
1.28 s are spent before the request reaches an engine. At each seed, \sys at equal quality
is 45--64\% and
65--72\% faster than these two.

\section{Tail latency}\label{app:tail}

Figure~\ref{fig:tail} plots quality against p99 E2E latency. Each point is filled or hollow by quality, mean latency
and cost, as in the main text's Figures~\ref{fig:frontier} and~\ref{fig:gemma}. Table~\ref{tab:tail} gives \sys's p99 at equal quality against each
family's baseline settings, computed as the main text's equal-quality comparison with p99 in place of the mean. It is lower against 237 of the 246 Qwen settings and 12 of the
15 Gemma settings it can be compared with, by a median of 37\% and
24\%.

\end{document}